# Metasurface interferometry towards quantum sensors

Philip Georgi[1*], Marcello Massaro[1*], Kai-Hong Luo[1*], Basudeb Sain[1], Nicola Montaut[1], Harald Herrmann[1], Thomas Weiss[2], Guixin Li[3], Christine Silberhorn[1], Thomas Zentgraf[1]

[1] Paderborn University, Department of Physics, Warburger Str. 100, 33098 Paderborn, Germany

[2] University of Stuttgart, 4th Physics Institute, Pfaffenwaldring 57, 70569 Stuttgart, Germany

[3] Southern University of Science and Technology, Department of Materials Science and Engineering & Shenzhen Institute for Quantum Science and Engineering, 518055 Shenzhen, China

## Abstract

**Optical metasurfaces open new avenues for precise wavefront control of light for integrated quantum technology. Here, we demonstrate a hybrid integrated quantum photonic system that is capable to entangle and disentangle two-photon spin states at a dielectric metasurface. By interfering single-photon pairs at a nanostructured dielectric metasurface, a path-entangled two-photon NOON state with circular polarization is generated that exhibits a quantum HOM interference visibility of 86 ± 4%. Furthermore, we demonstrate nonclassicality and phase sensitivity in a metasurface-based interferometer with a fringe visibility of 86.8 ± 1.1 % in the coincidence counts. This high visibility proves the metasurface-induced path entanglement inside the interferometer. Our findings provide a promising way to hybrid-integrated quantum technology with high-dimensional functionalities in various applications like imaging, sensing, and computing.**

## Introduction

Integrated quantum technology is widely used for enabling quantum applications in various systems for secure quantum communication [1, 2] as well as in quantum simulation [3, 4, 5, 6, 7] and quantum metrology [8]. Although significant progress of large-scale integration in one generic material has been achieved [9], highly miniaturized integrated systems are required for more complex functionalities, like teleporting of twisted photons [10] in a high-dimensional spin-orbital angular momentum (OAM) space [11]. However, the on-chip manipulation of circularly polarized photons is still an unsolved problem.

State of the art metasurfaces achieve essentially any kind of manipulation of light wavefronts for applications like ultra-flat lenses for imaging [12, 13], vector beam generation [14, 15], optical holography [16] and even nonlinear phase manipulation [17, 18]. While all these concepts solely rely on the classical electromagnetic description, the potential of metasurfaces for quantum applications is still widely unexplored. However, as versatile optical elements for locally altering the amplitude, phase, and polarization of light [19, 20, 21], metasurfaces can provide new functionalities to miniaturized quantum systems.

Recently, few initiatives have been taken to investigate the potential of metasurfaces in quantum optics. Jha et al. theoretically proposed that a metasurface can induce quantum interference between orthogonal radiative transition states of atoms [22] and quantum entanglement between two qubits [23]. Later, it was demonstrated that entanglement of spin and orbital angular momentum of a single photon can be generated via a metasurface [24], also, metasurface can provide a compact solution for quantum state reconstruction [25]. However, so far, there is only limited experimental evidence whether metasurfaces are suited for state manipulation in quantum optical experiments. If the current technology of metasurfaces can be directly applied to practical quantum applications, they can offer advanced solutions for quantum imaging [26, 27], sensing [28] and computing [29].

Here, we demonstrate entanglement and disentanglement of two-photon states using an all-dielectric metasurface. Our metasurface allows the generation of path-entangled NOON states with circular polarization due to the quantum interference effect. We observe a photon bunching within two spatially distinct output channels of the metasurface. Passing the same metasurface the second time, the generated path-entangled two-photon spin state can be disentangled, without introducing additional phase information. Our experiments indicate that metasurfaces are perfectly suited to provide large-scale and high-dimensional quantum functionalities with properties that go far beyond the conventional optical elements. Thus, hybrid integration of quantum optical elements together with metasurfaces offers the promise for delivering robust multi-photon entanglement and high-dimensional quantum applications [30].

## Results

**Metasurface design and functionality**

For the experiment, we designed the metasurface to deflect the incident light into two different output directions under angles of $\pm 10°$. The deflection is obtained by using a space-variant Pancharatnam-Berry-phase that results from the polarization conversion for the transmitted light [31]. As a platform for the metasurface, we use silicon nanofin structures. They act as local halfwave plates, which convert the circular polarization states into their cross polarization and add a spatial phase term based on their orientation angle. For our design, we choose a linear phase gradient, which diffracts the incoming light under the desired angle. We note that the sign of the phase gradient explicitly depends on the helicity of the circularly polarized light, which effectively makes our metasurface essentially a spatial separator for the circular polarization states (for more details see Supplementary Information).

**Entanglement and disentanglement**

Precise control and preparation of multi-photon entanglement are of fundamental interest for quantum technologies. With the accurate manipulation of the single photons' wavefront by a metasurface, spatial and polarization-based entanglement can be achieved. In our case, the metasurface is designed to spatially separate the generated circular polarization states of light and thus enforces a quantum state's representation in that particular basis, as sketched in Figure 1a. For the particular quantum state $|\Psi\rangle = \hat{a}_H^\dagger \hat{a}_V^\dagger |0\rangle$ such a process at the metasurface leads to an entangled state. Note, that we use the index $H$ and $V$ for the two linear polarizations states and $L$ and $R$ for the circular polarization states. Starting with relations between the photon creation operators

$$\hat{a}_L^\dagger = \frac{1}{\sqrt{2}} \left(\hat{a}_H^\dagger + i \cdot \hat{a}_V^\dagger\right)$$
$$\hat{a}_R^\dagger = \frac{1}{\sqrt{2}} \left(\hat{a}_H^\dagger - i \cdot \hat{a}_V^\dagger\right),$$
(1)

The metasurface enforces a change of the initial quantum state $|\Psi\rangle$ in the circular basis as

$$|\Psi\rangle = \hat{a}_H^\dagger \hat{a}_V^\dagger |0\rangle = -\frac{i}{2} \left(\hat{a}_L^\dagger \hat{a}_L^\dagger - \hat{a}_R^\dagger \hat{a}_R^\dagger\right) |0\rangle \quad (2)$$

which corresponds to a two-photon NOON state. Through the metasurface induced spatial separation, this path-encoded quantum state cannot be decomposed into two single-photon states and is therefore entangled.

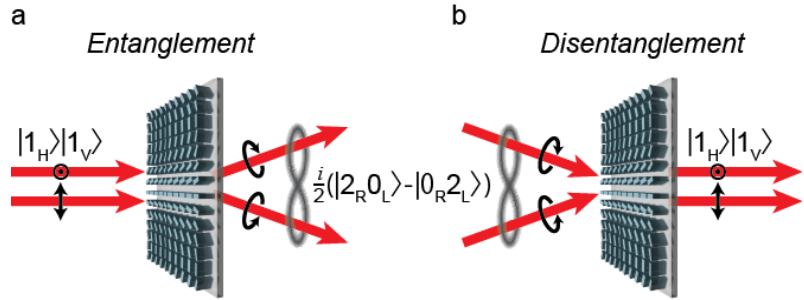

**Fig. 1. Spatial entanglement and disentanglement of a two-photon state at a metasurface. a,** If a single photon pair with orthogonal linear polarization is inserted into the metasurface, it is divided into its circular polarization components. Since the inserted quantum state is a two-photon NOON state in the circular polarization basis, both photons will always choose the same output channel and spatial entanglement is obtained. **b,** The situation inverses for the insertion of a spatially entangled NOON state into the metasurface. The quantum interference leads to a projection back to the original orthogonal linear polarization states and the photon pair is spatially disentangled.

Interestingly, the same metasurface can be used in the reverse process in which the circular polarization states are spatially recombined to disentangle the generated state. We utilize the polarization basis change functionality to build-up a metasurface-based interferometer. By introducing a phase delay $\varphi$ between the two circular polarization channels and use the conversion of the polarization states of the metasurface, we obtain the quantum state

$$|\Psi\rangle_\varphi = -\frac{i}{2} \left(\hat{a}_L^\dagger \hat{a}_L^\dagger e^{-i2\varphi} - \hat{a}_R^\dagger \hat{a}_R^\dagger\right) |0\rangle$$
$$= \frac{1}{2} \left(e^{-i2\varphi} + 1\right) \hat{a}_H^\dagger \hat{a}_V^\dagger |0\rangle - \frac{i}{4} \left(e^{-i2\varphi} - 1\right)(\hat{a}_H^\dagger \hat{a}_H^\dagger - \hat{a}_V^\dagger \hat{a}_V^\dagger) |0\rangle$$
(3)

For the case that $\varphi = n\pi, (n\epsilon\mathbb{Z})$, the output state after passing the metasurface twice is the same as the initial two-photon state $\hat{a}_H^\dagger \hat{a}_V^\dagger |0\rangle$, which is disentangled (as shown in Figure 1b). In the case of $\varphi = \left(n + \frac{1}{2}\right)\pi, (n\epsilon\mathbb{Z})$, the output state $i \cdot \frac{1}{2} \left(\hat{a}_H^\dagger \hat{a}_H^\dagger - \hat{a}_V^\dagger \hat{a}_V^\dagger\right) |0\rangle$ is entangled even after spatial recombination, as it cannot be decomposed in neither polarization base.

**Generation of NOON spin states**

First, we investigate the NOON-state generation at the metasurface. A key quantum feature of the NOON state is its photon bunching characteristics, i.e. we expect that both photons will always choose the same metasurface output channel after inserting the quantum state $\hat{a}_H^\dagger \hat{a}_V^\dagger |0\rangle$. To demonstrate this bunching effect, we use the setup shown in Figure 2. It contains four logical parts: a two-photon source for the generation of the initial quantum state $|1_H\rangle|1_V\rangle$, a Michelson interferometer to adjust the time delay between these two photons, a metasurface as a quantum interference device, and a single photon detection system. To control the time delay between the two photons, we use a modified Michelson interferometer with a polarizing beam splitter cube and two quarter-wave plates. Since the time delay directly influences the temporal overlap between both photons, we utilize it to enable ($\tau = 0$) and disable ($\tau \to \infty$) the quantum interference effect at the metasurface. In the case of no interference, the two photons will choose either output with a 50% chance, while in the case of perfect interference both photons will always choose the same random output port.

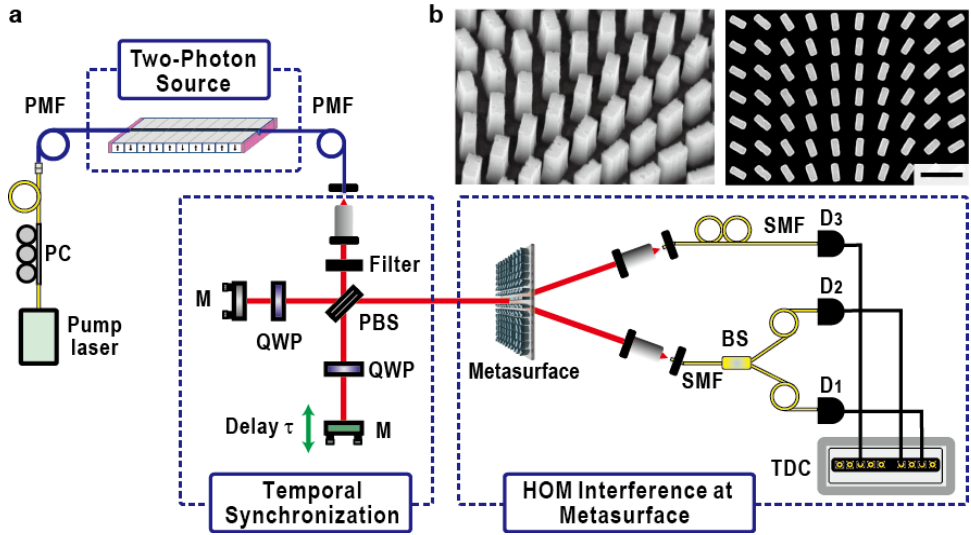

**Fig. 2. Illustration of the measurement setup. a,** The two-photon source creates a photon pair with orthogonal linear polarizations by spontaneous parametric down-conversion (SPDC). The two photons are temporally delayed relative to each other by $\tau$ with a Michelson interferometer, containing a polarizing beam splitter (PBS) and quarter-wave plates (QWPs). The photon pair passes through the metasurface where the NOON state is generated. The spatially entangled state is then analyzed by a coincidence measurement system in two different configurations with single photon detectors ($D_n$). (PMF – polarization maintaining fiber, SMF – single mode fiber, M – mirror, PC – polarization controller, TDC – time-digital converter) **b,** Scanning electron microscopy images under 45° (left) and top view (right) for a small area of the fabricated silicon metasurface (scale bar 1 µm).

To characterize the generated quantum state, we perform two different coincidence measurements for various time delays $\tau$. First, we measure the number of coincidences between both outputs channels of the metasurface. The generated NOON state should not contribute to the number of registered coincidences, ideally leading to zero coincidence counts. However, when the initial $|1_H\rangle$ and $|1_V\rangle$ photons do not arrive simultaneously at the metasurface, we expect a 50% coincidence probability per inserted photon pair (without losses). Thus, as we vary the time delay $\tau$, we expect the well-known Hong-Ou-Mandel (HOM) correlation dip. Note that a visibility higher than 50% verifies the quantum character of interference for such HOM experiments [32]. Second, we measure the number of coincidences between the two outputs of a 50:50 beam splitter that has been placed in one of the output channels of the metasurface. Coincidence counts for this measurement can only be obtained when both photons choose the same output channel of the metasurface. Since the probability for this event is twice as large in the case of interference, we expect a 2:1 ratio in the coincidence counts between $\tau = 0$ and $\tau \to \infty$. Both coincidence measurements are performed simultaneously by using three superconducting nanowire single-photon detectors (SNSPDs). One of these detectors ($D_3$) is directly connected to one of the metasurface output channels, while the other two detectors ($D_1$ and $D_2$) are placed behind an integrated beam splitter (a 3-dB fiber coupler), which is connected to the second metasurface output channel. In this configuration, the first coincidence measurement between the two metasurface channels can be calculated as $C_{13} + C_{23}$, where $C_{ij}$ refers to the coincidence between the detectors $D_i$ and $D_j$.

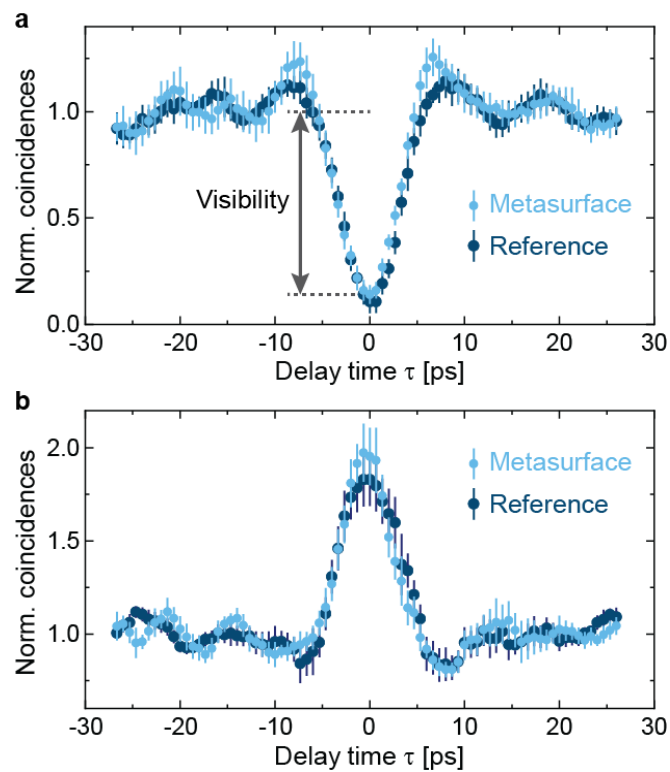

**Fig. 3. Hong-Ou-Mandel coincidence measurements. a,** Normalized coincidence counts ($C_{13} + C_{23}$) between the two output channels of the metasurface for a variation of the initial photon time delay. The high visibility beyond the classical limit of 50% confirms the expected quantum interference effect. The reference is obtained for a standard Hong-Ou-Mandel experiment with a standard beam splitter to characterize the quality of the photon source. **b,** Normalized coincidence counts $C_{12}$ between detectors $D_1$ and $D_2$ for the measurement with the polarization beam splitter in one output channel of the metasurface. The peak in the coincidence counts of the "anti"-HOM measurement confirms that the two photons always take the same output channel.

The coincidences show a clear dip at zero-time delay with high visibility of $86 \pm 4\ \%$ (Figure 3a). Note that the visibility clearly exceeds the limit of 50%, which can be achieved with classical coherent light.

At the same time, the second coincidence measurement ($C_{12}$) shows at $\tau = 0$ a clear coincidence peak ("anti"-HOM peak), which confirms that the two photons are always bunched together in one output channel (Figure 3b). In case of interference, both measurements show that the probability of at least one photon in each output channels decreases (first measurement), while the probability of at least two photons in one output channel increases (second measurement). Thus, photon bunching occurs (for details see Supplementary Information).

**Metasurface-based interferometer**

Next, we study whether the metasurface is also preserving the quantum coherence of the generated state, i.e. the phase relations between the components of the superposition are fixed and not randomly redistributed. Such coherence is important in quantum metrology applications, where phase-measurements play a key role. For that, we realized a folded metasurface-based interferometer (MBI) in which the photons pass the metasurface twice (Figure 4a). By tilting a 130-µm-thick glass plate in one of the two arms of the interferometer, we introduce a phase difference between the two optical paths. The final state is then separated and analyzed at a polarizing beam splitter (PBS).

First, we use a strongly attenuated laser as a weak coherent input state. For the input state $|\alpha\rangle_H |0\rangle_V$, i.e. a coherent state in one input and vacuum in the other, the count rates in both output ports of the MBI show an oscillating behavior. At the same time, the coincidences show an oscillating behavior with the doubled frequency following a $sin^2(\varphi)$ function (for details see Figure S2b of the Supplementary Information). This behavior can be understood from a classical point of view, where the coincidence corresponds to the product of the single counts. The fringe contrast of $90 \pm 1$ % is in good agreement with our theoretical estimations (see Supplementary Information). Note that $\varphi$ corresponds to the introduced total phase by passing twice through the tilted glass plate.

Next, we launch one photon per input mode $|1\rangle_H |1\rangle_V$ by using the photon pairs from the SPDC-source. We observe that the counts at each individual MBI output channel will be constant regardless of the introduced phase $\varphi$ (Figure 4c). This is due to the first-order correlation of two MBI outputs being independent of $\varphi$ (see Supplementary Information). However, when we determine the coincidences between the two outputs, we observe the same double frequency oscillation from the coherent case, which is now phase shifted following a $cos^2(\varphi)$ function. When two orthogonally polarized photons arrive at the metasurface simultaneously ($\tau = 0$), there is no coincidence contribution in the Hong-Ou-Mandel experiment due to the photon bunching effect (as shown in Figure 4b). In this scenario, path-entangled photon pairs are generated at the metasurface. Thus, the visibility of interference fringes from the MBI is $86.8 \pm 1.1$ %, which is beyond the violation limit of Bell's inequalities (70.7%). Besides, additional experiments for various time-delays $\tau$ between the initial two photons show a reduced visibility of the coincidence rate (see Figure 4c). When one photon is delayed by 3 ps, the normalized HOM coincidence rate of around 50% tells us that the two photons still overlap partially in time. For this partial overlap, the visibility of the interference fringes is $67 \pm 2$ %, which is close to the boundary of the Bell's inequalities. If one photon is delayed more than 17 ps, they arrive at the MBI one after the other. Correspondingly, there is no path-entanglement generated inside the MBI. This is in good agreement with our experimental visibility of 45 ± 5 % for the coincidence rate.

In contrast to the coherent case, the coincidence counts can no longer be perceived as the product of the single counts. This pure nonclassical effect is a key feature of quantum interferometry that is closely related to photon entanglement. The peaks in the coincidences result from the second-order correlation of the entangled NOON spin state with circular polarization, which is generated after passing the metasurface the first time. Passing the entire MBI, the entangled NOON spin state is disentangled into the original two-photon state $\hat{a}_H^\dagger \hat{a}_V^\dagger |0\rangle$, if there is no phase difference between two MBI arms (first term in Eq. (3)). After splitting the state at the PBS, the two photons arrive at the two detectors simultaneously, which causes the maximum coincidence rates. On the other hand, the minima in the coincidences result from the second-order correlation of the entangled NOON state with linear polarizations (second term in Eq. (3)). For phase differences of $\varphi = \left(n + \frac{1}{2}\right)\pi$, $(n \in \mathbb{Z})$, the output of MBI is solely determined by the second term of Eq. (3). The two photons are either in the H or V path, but they cannot be in both paths simultaneously. Hence, there is no coincidence contribution, similarly to the HOM-effect as shown in Figure 3.

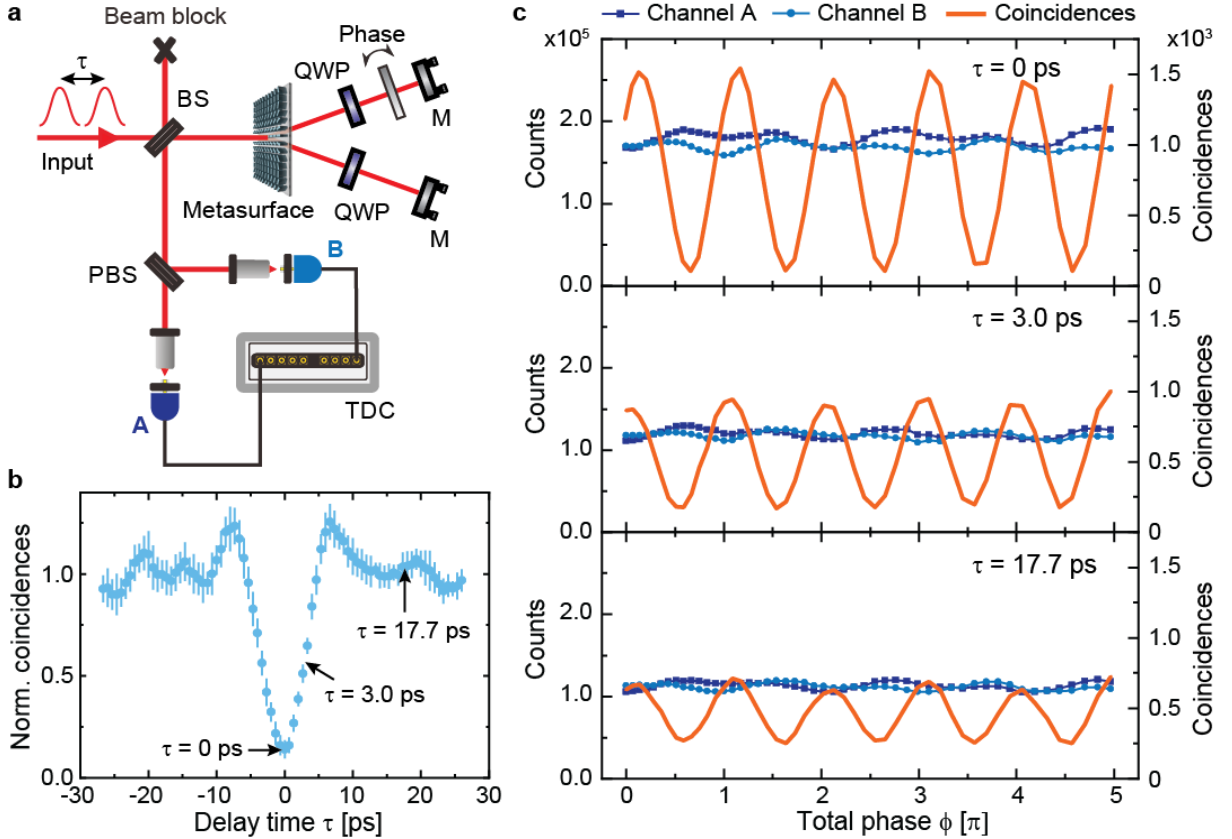

**Figure 4. a,** Schematic view of the metasurface-based interferometer using a time-delay τ between the two input photons. **b,** Selected time-delays τ illustrated at the HOM-dip measurement. **c,** Experimental results for the two-photon state with three different time-delays τ after passing through the MBI. The count rates at both detectors are independent of the introduced phase φ and the time-delay τ. The measured coincidence rates show an oscillatory

behavior with lower visibility at higher delay-times. The obtained visibility values are (86.8±1.1)% for the case of no time delay, (67±2)% for $\tau = 3.0$ ps and (44±5)% for $\tau = 17.7$ ps.

## Discussion

For quantifying the quality of the quantum interference and the entanglement that takes place at the metasurface, we compare the visibility of the HOM dip obtained for the metasurface with a reference experiment. The reference experiment is inspired by Grice and Walmsley [32] and allows us to determine an upper bound on the achievable HOM visibility, which is around $89 \pm 5$ % (for more details see Supplementary Information). The visibility of the HOM dip, which depends not only on the quality of the SPDC source but also on the interference strength of the used interference device, clearly surpasses the classical threshold of 50%. Hence, the high HOM visibility of $86 \pm 4$ % shows the quantum mechanical nature of the measured spatially entangled two-photon NOON state generated at the metasurface.

The phase information inside the MBI is hidden when doing an intensity (first-order correlation) measurement at each detector. This phase information can be revealed with the help of coincidence (second-order correlation) measurements between the two detectors. Such a result is similar to the Franson-interferometer that is used for verifying the energy-time entanglement [33] and for security information coding in quantum cryptography. Furthermore, our metasurface-based interferometer shows the feasibility of quantum sensors based on nanostructured metasurfaces. Note, that the double period of the fringes itself is not a quantum signature of the two-photon NOON state since it also appears in the coincidence measurement for the weak coherent state.[34]

Integrated photonic quantum experiments are routinely performed using large-scale optical components such as directional couplers and beam splitters. With our experiments, we demonstrate the preparation of spatial entanglement and disentanglement based on a metasurface in a more compact setting. The results are especially remarkable since the metasurface consists of spatially distributed nanostructured elements with slightly different scattering properties. The experiments confirm that quantum entanglement and interference take place at our dielectric metasurface while phase sensitive effects (quantum coherence) are preserved. Our findings demonstrate that metasurfaces can achieve a similar interference performance as traditional optical components and are indeed viable candidates for integrated quantum nano-sensors and quantum interferometry.

Here, we focused on a metasurface for entanglement and interference and therefore for state manipulation purposes. Our metasurface operates as one of the basic building blocks of typical photonic quantum circuits, which splits and recombines optical modes in nested interferometers. However, metasurfaces have enormous potential in quantum optics. Their ability to fully control the wavefronts of light can be used to generate multi-photon and high-dimensional entanglement with different spin-OAM. Combining multiple optical functionalities into a single metasurface as an efficient and compact quantum optical device might dramatically improve the performance and even lead to new concepts for practical quantum applications. Since metasurfaces can be used directly at waveguides and fiber-end facets, such hybrid nanophotonics systems for arbitrary basis transformation can be used for robust integrated quantum technologies, from sensor arrays to quantum simulators. In this context, future research has to show whether metasurfaces can directly generate quantum states, like single photon pairs, in a well-defined and efficient way without the need for an additional source.

**Acknowledgments:** The authors would like to acknowledge support from Cedrik Meier (University of Paderborn) by providing his electron beam lithography system and Fabian Meier for assisting with the fabrication. This project has received funding from the European Research Council (ERC) under the European Union's Horizon 2020 research and innovation programme (grant agreement No. 724306) and the Deutsche Forschungsgemeinschaft (DFG, German Research Foundation) through the Collaborated Research Center TRR 142 (No. 231447078). G.L. is supported by the National Natural Science Foundation of China (Grant No. 11774145), Applied Science and Technology Project of Guangdong Science and Technology Department (2017B090918001).

**Conflict of interests:** The authors declare that they have no conflict of interest.

**Data availability:** All data needed to evaluate the conclusions in the paper are present in the paper and/or the Supplementary Information. Additional data related to this paper may be requested from the authors.

**Author contributions:** P.G., M.M., and K.H.L. contributed equally to this work. P.G., M.M., and K.H.L. performed the experiments, K.H.L. and G.L. proposed the idea, P.G. and T.W. conducted the designs and numerical simulations, B.S. fabricated the sample, N.M and H.H. prepared the quantum fiber-pigtailed source. C.S. and T.Z. supervised the overall project. All the authors analyzed the data, discussed the results and prepared the manuscript.